# Categorical data analysis using discretization of continuous variables to investigate associations in marine ecosystems

Hiroko Kato Solvang [*]

Institute of Marine Research, PO Box 1870 Nordnes, N-5817 Bergen, Norway
hiroko.solvang@hi.no

Shinpei Imori
Hiroshima University, 1-4-1 Kagamiyama, Higashi-Hiroshima City, Hiroshima, 739-8527, Japan
imori@hiroshima-u.ac.jp

Martin Biuw
Institute of Marine Research, Fram Centre, PO Box 6606 Langnes, N-9296 Tromsø, Norway
martin.biuw@hi.no

Ulf Lindstrøm
Institute of Marine Research, Fram Centre, PO Box 6606 Langnes, N-9296 Tromsø, Norway
Arctic University of Norway, 9296 Tromsø, Norway
ulf.lindstroem@hi.no

Tore Haug
Institute of Marine Research, Fram Centre, PO Box 6606 Langnes, N-9296 Tromsø, Norway
tore.haug@hi.no

* Corresponding author

Declarations

Data availability: The data we applied in Real-data analysis is available on request.

Funding statement: This work was supported by JSPS Bilateral Program, Japan Grant Number JPJSBP120219927.

Conflict of interest: The authors declare that they have no conflict of interest.

Ethical approval: All applicable international, national, and/or institutional guidelines for the care and use of animals were followed.

Abstract

Understanding and predicting interactions between predators and prey and their environment are fundamental for understanding food web structure, dynamics, and ecosystem function in both terrestrial and marine ecosystems. Thus, estimating the conditional associations between species and their environments is important for exploring connections or cooperative links in the ecosystem, which in turn can help to clarify such causal relationships. For this purpose, a relevant and practical statistical method is required to link presence/absence observations with biomass, abundance, and physical quantities obtained as continuous real values. These data are sometimes sparse in oceanic space and too short as time series data. To meet this challenge, we provide an approach based on applying categorical data analysis to present/absent observations and real-number data. This approach consists of a two-step procedure for categorical data analysis: 1) finding the appropriate threshold to discretize the real-number data for applying an independent test; and 2) identifying the best conditional probability model to investigate the possible associations among the data based on a statistical information criterion. We conduct a simulation study to validate our proposed approach. Furthermore, the approach is applied to two datasets: 1) one collected during an international synoptic krill survey in the Scotia Sea west of the Antarctic Peninsula to investigate associations among krill, fin whale (*Balaenoptera physalus*), surface temperature, depth, slope in depth (flatter or steeper terrain), and temperature gradient (slope in temperature); 2) the other collected by ecosystem surveys conducted during August-September in 2014–2017 to investigate associations among common minke whales, the predatory fish Atlantic cod, and their main prey groups (zooplankton, 0-group fish) in Arctic Ocean waters to the west and north of Svalbard, Norway. The R code summarizing our proposed numerical procedure is presented in Supplementary file.

1. Introduction

Recent climate change and fluctuations in the environmental conditions of oceans have affected the spatial distribution of both zooplankton and fish species known to be the prey of marine mammals. Investigating the associations among biological communities and oceanographic factors is important for exploring connections or cooperative links in the ecosystem, which may in turn further clarify these relationships. For marine ecosystem assessment, several spatial and temporal data are obtained by oceanographic surveys to investigate the changes occurring due to global climate change or human activity. However, these data can be either too sparsely or too densely distributed, depending on the area in the ocean, for application to a spatial model that assumes the data are distributed uniformly, and as time series data they are much too short as annual observations for application to a conventional series model that considers temporal correlations. For spatially sparse data, several statistical spatial models have been developed (e.g., Smith et al., 2014; Ver Hoef and Jansen, 2007; Sugasawa et al., 2022), and an integrated approach to investigating the trends of short time series data was also proposed (Solvang and Planque, 2020). In addition to such efforts, Solvang et al. (2021) proposed a practical approach to analyzing the data as categorical data for investigating the relationships among common minke whales (*Balaenoptera acutorostrata*) and their prey species using spatial and temporal data. This approach is useful because it is not necessary to consider the statistical properties of spatial sparseness or short time series for the observations in advance. The original idea of our approach derives from Sakamoto and Akaike (1979), which was the first study to apply the Akaike Information Criterion (AIC, see Akaike, 1974) to find the best association for the response variable from possible explanatory variables. The computational calculation procedure was first developed in Fortran code and called CATDAP (Katsura and Sakamoto, 1980). Now, the code has been edited in R (2023) and implemented as '*catdap*' in the R package (2023). The fist computational procedure was available for only categorical data; however, *catdap* in R is able to handle data including continuous variables by applying a histogram model to the continuous variables, and it tries to find the best categorical groups by AIC. The procedure is automatically conducted without any prior information, and the continuous data are sometimes categorized into many groups. While the outputs may numerically make sense, the interpretation of many categorized groups becomes too difficult for experts, especially those who investigate biological and ecological associations in marine ecosystems. For them, it might be useful if the data were simply categorized as binary to interpret the data as, e.g., higher/lower or presence/absence, according to a single threshold. In this study, we propose a useful statistical method to find the optimum threshold to simply categorize the continuous data into two groups and present binary data by considering the relationships among the data based on the procedure in CATDAP. This threshold presents the best association between the response variable and the explanatory variable. Based on the proposed procedure, each continuous data item is replaced by two categorical data. Then, we integrate the data as one dataset and finally apply *catdap* to the dataset to find the best causal relationships among several combinations through setting a response variable and explanatory variables for the data.

This article is organized as follows. Section 2 reviews the categorical data analysis based on AIC from Sakamoto and Akaike (1979) and introduces our proposed method. The simulation study to verify the proposed method is presented in Section 3. Section 4 gives two examples for the real-data analysis—one is applied to the dataset collected by a combined krill and whale survey in the Scotia Sea in the southwest Atlantic sector of the Southern Ocean in 2019 (Macaulay et al., 2019; Krafft et al.. 2019, 2021), and the other is applied to the dataset collected by ecosystem surveys conducted during August-September in 2014–2017 to investigate associations among common minke whales, the predatory fish

Atlantic cod, and their main prey groups (zooplankton, 0-group fish) in Arctic Ocean waters to the west and north of Svalbard, Norway (Solvang et al., 2021).

2. Method

2.1. Independent test using AIC evaluation in the approach of Sakamoto and Akaike (1978)

This subsection reviews the categorical data analysis based on AIC provided by Sakamoto and Akaike (1978). We examine the method called CATDAP. For easy understanding, the dataset we take as an example includes four types of data: whale counting number, krill biomass, depth in the sea, and surface temperature of the sea. Let us consider a simple relationship between whale and krill. For the categorical data analysis, a contingency table is set by counting the numbers for two categories, presence and absence, from the data of whale and krill.

Let $W, K \in \{0,1\}$ be binary random variables. Suppose that we observe $n$ samples $(W_i, K_i)$ $(i = 1, \cdots, n)$ for whale counting number and krill biomass, which are independently and identically distributed with the same distribution as $(W, K)$. For $j, k = 0, 1$, let $n(j,k) = \#\{(W_i, K_i) = (j,k), i = 1, \cdots, n\}$, $n(j,\bullet) = n(j,0) + n(j,1)$, and $n(\bullet, k) = n(0,k) + n(1,k)$. Then, we construct a two-way contingency table as below:

|  | $W = 1$ | $W = 0$ | Total |
|---|---|---|---|
| $K = 1$ | $n(1,1)$ | $n(1,0)$ | $n(1,\bullet)$ |
| $K = 0$ | $n(0,1)$ | $n(0,0)$ | $n(0,\bullet)$ |
| Total | $n(\bullet,1)$ | $n(\bullet,0)$ | $n$ |

Here, $W = 1$ and $K = 1$ mean presence of whale and presence of krill, and $W = 0$ and $K = 0$ mean absence of whale and absence of krill. Let the cell frequency and joint probability for whale and krill be represented by $n(i_w, i_k)$ $(i_w = 0,1, \ i_k = 0,1)$ and $p(i_w, i_k)$, where $\sum_{i_w=0}^{1} \sum_{i_k=0}^{1} n(i_w, i_k) = n$ and $\sum_{i_w=0}^{1} \sum_{i_k=0}^{1} p(i_w, i_k) = 1$. Assuming that $n(i_w, i_k)$ has a multinomial distribution with unknown probabilities $p(i_w, i_k)$, the probability is given by

$$P(\{n(i_w, i_k)\} \mid \{p(i_w, i_k)\}) = \frac{n!}{\prod_{i_w=0}^{1} \prod_{i_k=0}^{1} n(i_w, i_k)} \prod_{i_w=0}^{1} \prod_{i_k=0}^{1} p(i_w, i_k)^{n(i_w, i_k)}. \quad (1)$$

If the logarithm of the first term $\dfrac{n!}{\prod_{i_w=0}^{1}\prod_{i_k=0}^{1} n(i_w,i_k)}$ of (1) is replaced by $M$, the log-likelihood of the unknown parameter is given by

$$l(\{p(i_w,i_k)\}) = M + \sum_{i_w=1}^{1}\sum_{i_k=0}^{1} n(i_w,i_k)\log p(i_w,i_k). \tag{2}$$

If it is assumed that whale presence is independent of krill presence, the model is described by

$$\text{Model(i): } p(i_w,i_k) = \theta(i_w,\bullet)\theta(\bullet,i_k),$$

where $\theta(i_w,\bullet) = \sum_{i_k=0}^{1} p(i_w,i_k)$ and $\theta(\bullet,i_k) = \sum_{i_w=0}^{1} p(i_w,i_k)$. The log-likelihood is represented by

$$l(\{\theta(i_w,\bullet),\theta(\bullet,i_k)\}) = M + \sum_{i_w=0}^{1}\sum_{i_k=0}^{1} n(i_w,i_k)\log \theta(i_w,\bullet)\,\theta(\bullet,i_k), \tag{3}$$

where $\sum_{i_w=0}^{1}\theta(i_w,\bullet) = \sum_{i_k=0}^{1}\theta(\bullet,i_k) = 1$. The maximum likelihood estimators are obtained by maximizing Eq. (3) as follows:

$$\hat{\theta}(i_w,\bullet) = \frac{n(i_w,\bullet)}{n} \text{ and } \hat{\theta}(\bullet,i_k) = \frac{n(\bullet,i_k)}{n}.$$

On the other hand, if it is assumed that whale presence depends on krill presence, the model is described by

$$\text{Model(d): } p(i_w,i_k) = \theta(i_w,i_k),$$

where $\sum_{i_w=0}^{1}\sum_{i_k=0}^{1} \theta(i_w,i_k) = 1$. The log-likelihood is given by

$$l(\{\theta(i_w,i_k)\}) = M + \sum_{i_w=0}^{1}\sum_{i_k=0}^{1} n(i_w,i_k)\log \theta(i_w,i_k). \tag{4}$$

The maximum likelihood estimators are given by $\hat{\theta}(i_w,i_k) = \dfrac{n(i_w,i_k)}{n}$. The comparison of Model(i) with Model(d) is usually done by Person's Chi-square test of independence (Pearson, 1990). The Chi-square statistics is calculated by the disparity between the expected value and observation and then assessed by a Chi-square distribution with 1 degree of freedom for a null hypothesis supporting Model(i). Sakamoto and Akaike (1974) used a statistical model selection approach from the model candidates Model(i) and Model(d) in CATDAP, without using the Chi-square independence test-based null hypothesis. The discrepancy of a model fitted to a set of observed data by maximum likelihood is evaluated by AIC, given by -2 × log (maximized likelihood) + 2 × number of parameters in the model. AICs for Model(i) and Model(d) are given by

$$\text{AIC}_{\text{Model(i)}} = -2\left\{M + \sum\nolimits_{i_w=0}^{1}\sum\nolimits_{i_k=0}^{1} n(i_w, i_k)\log\frac{n(i_w,\bullet)n(\bullet,i_k)}{n^2}\right\} + 2\{(2-1)+(2-1)\} \tag{5}$$

and

$$\text{AIC}_{\text{Model(d)}} = -2\left\{M + \sum\nolimits_{i_w=0}^{1}\sum\nolimits_{i_k=0}^{1} n(i_w, i_k)\log\frac{n(i_w, i_k)}{n}\right\} + 2(2\times 2-1). \tag{6}$$

To compare the fitting of the two models to the data, $\Delta\text{AIC} = \text{AIC}_{\text{Model(i)}} - \text{AIC}_{\text{Model(d)}}$ is calculated. If $\Delta\text{AIC} < 0$, Model(i) should be adopted, that is, the presence of whale and krill is independent, otherwise Model(d) should be adopted, which means that the presence of whale is dependent on the presence of krill.

The above example can be expanded to an $m$-way contingency table that consists of a variable to be predicted and the $m-1$ predictors. When considering the $m-1$ explanatory variables for the responsive variable $I_1$ in general, the $2^{m-1}$ ($= {}_{m-1}C_{m-1} + \cdots + {}_{m-1}C_0$) models to compare the combinations of all explanatory variables are given by

$$\begin{aligned}
&\text{Model } (I_1; I_2, \cdots, I_m) : p(i_1 | i_2, \cdots, i_m) = \theta(i_1 | i_2, \cdots, i_m), \\
&\text{Model } (I_1; I_2, \cdots, I_{m-1}) : p(i_1 | i_2, \cdots, i_m) = \theta(i_1 | i_2, \cdots, i_{m-1}), \\
&\qquad\qquad\cdots\cdots\cdots \\
&\text{Model } (I_1; I_3, \cdots, I_m) : p(i_1 | i_2, \cdots, i_m) = \theta(i_1 | i_3, \cdots, i_m), \\
&\qquad\qquad\vdots \qquad\qquad\qquad\vdots \\
&\text{Model } (I_1; I_m) : \qquad p(i_1 | i_2, \cdots, i_m) = \theta(i_1 | i_m), \\
&\qquad\qquad\cdots\cdots\cdots \\
&\text{Model } (I_1; I_2) : \qquad p(i_1 | i_2, \cdots, i_m) = \theta(i_1 | i_2), \text{ and} \\
&\text{Model } (I_1; \phi) : \qquad p(i_1 | i_2, \cdots, i_m) = \theta(i_1).
\end{aligned} \tag{7}$$

The above models are summarized by

$$\text{Model } (I_1; \mathbf{J}) : \qquad p(i_1 | \mathbf{i}) = \theta(i_1 | \mathbf{j}) , \tag{8}$$

where $\mathbf{I}$ is defined by the set of the explanatory variables $\{I_2, \cdots, I_m\}$, $\mathbf{J}$ indicates the arbitrary subset, and $\mathbf{i}$ and $\mathbf{j}$ are represented by the realization of $\mathbf{I}$ and $\mathbf{J}$. Therefore, the AIC of Model $(I_1; \mathbf{J})$ is given by

$$\text{AIC}_{\text{Model }(I_1; \mathbf{J})} = -2\sum\nolimits_{i_1, \mathbf{j}} n(i_1, \mathbf{j})\log\frac{n \cdot n(i_1, \mathbf{j})}{n(i_1)n(\mathbf{j})} + 2(c_1 - 1)(c_\mathbf{J} - 1) , \tag{9}$$

where $c_1$, $n(\mathbf{j})$, and $c_\mathbf{J}$ indicate the number of category of $i_1$, the marginal frequency for each combination of explanatory variables, and the number of category for $\mathbf{J}$. AIC is used to search for the optimal predictor on which the variable to be predicted has the strongest dependence (Sakamoto and Akaike, 1978). The procedure in CATDAP is carried out to arrange possible combinations for the response variable and calculate AICs for the models (by running '*catdap2*' function with '*additional.output*' by listing all combinations of the considered model in the R package *catdap*). It is useful to search for the optimal combination of variables for the response variable to show a reasonable relationship among variables.

2.2 Categorization using histogram model in CATDAP

The data of whale abundance are recorded in sighting surveys and can be easily divided into presence for observed whales (i.e., data are over zero) and absence for no observation (zero counted data); however, the data of krill abundance are usually represented by a real number. In this case, a method to estimate the threshold for dividing such real numbers into categorical groups is required for categorical data analysis. In CATDAP, a histogram model for the data distribution shape is fitted to the real numbers, and AIC evaluates the best subset obtained by dividing the histogram's bins between the minimum and maximum values of the data. Let the section width of the histogram be indicated by $d$, given by $d = \frac{x_{max} - x_{min}}{m-1}$, where $x_{min}$ and $x_{max}$ are the minimum and maximum values of the observation and $m$ is the bin used for dividing the histogram. The frequency distribution of the histogram is assumed to follow a multinomial distribution. Let the frequency and the corresponding probability for each section of the histogram be $n(i)$ and $p(i)$ $(i=1,\cdots,c)$, where $c$ is the number of the section. The probability given a set of frequencies $\{n(i)\}$ is represented by $P(\{n(i)\}|\{p(i)\}) = \frac{n!}{\prod_{i=1}^{c} n(i)!} \prod_{i=1}^{c} p(i)^{n(i)}$.

Therefore, the log-likelihood and AIC for the histogram model fitted to the real numbers are given by

$$l(\{p(i)\}) = \log \frac{n!}{\prod_{i=1}^{c} n(i)!} + \sum_{i=1}^{c} n(i) \log p(i)$$ and

$$\text{AIC} = -2\left[ \log \frac{n!}{\prod_{i=1}^{c} n(i)!} + \sum_{i=1}^{c} n(i) \log p(i) \right] + 2\{(c-1) + N \log d\},$$ respectively. If the explanatory variable in CATDAP is in real numbers, this model fitting is applied to the data for the optimum categorization. Furthermore, CATDAP is applicable for the case of objective variables indicating real numbers.

2.3. Proposed method: discretization of real-number variable into binary

CATDAP explores the optimum number for categorization of real-number variables, and the categorized number is sometimes identified as more than two by AIC. For example, if temperature is categorized into four variables by minimum AIC, it may be difficult for oceanographic or ecological experts to interpret the meaning of the four categorical groups in a marine ecosystem. Categorization such as high/low temperature groups makes it simpler to interpret the output than using four categorized temperature groups. To achieve this simplified approach, we propose the following procedure to estimate the optimum threshold for two-value discretization of continuous variables. The word 'discretization' of real-number variables corresponds to the word 'categorization' of continuous variables into two variables.

Note that we observe a continuous random variable $K^c \in \mathbb{R}$ instead of $K$ in subsection 2.1. This means that we observe $n$ samples $(W_i, K_i^c)(i=1,\cdots,n)$ that are independently and identically distributed with the same distribution as $(W, K^c)$. When categorizing $K^c$ in a certain way, we can apply the method described in subsection 2.1 to the discretized data. If the discretization were applied to all of the observed data, the distribution of discretized samples would be complicated, that is, the assumption for i.i.d. is not supported and the same sample could not be used for evaluation of the models for conditional dependency or independency using the procedure in subsection 2.1. In consideration of this point, we randomly separate the dataset into two parts, say $G_1 = \{(W_i, K_i^c), i=1,\cdots,m\}$ and $G_2 = \{(W_i, K_i^c), i=m+1,\cdots,n\}$ with an integer $m$ $(1<m<n)$. Note that $G_1$ and $G_2$ are independent. Our procedure consists of three steps: (i) discretize $K^c$ based on $G_1$; (ii) select the best model by using $G_2$ where $K_i^c$ $(i=m+1,\cdots,n)$ is discretized in the manner of step (i); and, (iii) repeat steps (i) and (ii) 1000 times because the selection result depends on data separation. Then, by comparing the averaged criteria for model selection, we select the model with the lower value as the best one. In the following, we explain each step:

Step (i): data discretization

Suppose that $K^c \in [a,b]$, where $a$ and $b$ are constants satisfying $a<b$. Let $s_1,\cdots,s_L$ be predetermined $L$ threshold points between $a$ and $b$, i.e., $a<s_1<\cdots<s_L<b$. Let us fix $l \in \{1,\cdots,L\}$. Define $K_i(s) = I\{K_i^c < s_l\}$ for $i=1,\cdots,m$, where $I\{\bullet\}$ is an indicator function. Then, we can obtain a contingency table based on $\{(W_i, K_i(s)), i=1,\cdots,m\}$. It follows from Pearson's Chi-square independent test that a p-value is calculated for each $l = 1, \cdots, L$. We use $s_l$ with the smallest p-value among $l = 1, \cdots, L$.

Step (ii): model selection

Let $\hat{s}$ be the best threshold point selected in step (i). Let $n'_s(j,k) = \#\{(W_i, K_i(s)) = (j,k), i = m+1, \cdots, n\}$. Note that given $s$, $\mathbf{n}'_s = (n'_s(1,1), n'_s(1,0), n'_s(0,1), n'_s(0,0))$ follows a multinomial distribution with parameters $n - m$ and $\mathbf{p}'_s = (p'_s(1,1), p'_s(1,0), p'_s(0,1), p'_s(0,0))$, where $p'_s(j,k) = P(W, K(s))$ $(j, k = 0,1)$. Based on $D'_2 = \{(W_i, K_i(\hat{s})), i = m+1, \cdots, n\}$, the procedure introduced in 2.1 is applied to obtain AIC, given $\hat{s}$ for the conditional probability models. In this case, we set $m = [n/2]$, where $[x]$ denotes an integer part of $x > 0$.

Finally, the averaged values for the optimum threshold and AIC are calculated after repeating steps (i) and (ii) for 1000 times. The averaged AIC is used to evaluate which model is better among the model candidates. The optimum threshold for the best model is used as a final optimum threshold to divide the continuous values into two categorical data, 1 and 0. If other continuous data like temperature or depth should be considered, corresponding to the counted whale number data, the same procedure is applied to two partitions for the continuous data. Here, we observe $n$ samples $(W_i, T_i^c)$ and $(W_i, D_i^c)$ $(i = 1, \cdots, n)$ for temperature ($T^c$) and depth ($D^c$) in addition to krill, which are independently and identically distributed with the same distributions as $(W, T^c)$ and $(W, D^c)$. Similar to the previous example for whale and krill, the two are also randomly separated based on the same integer $m$ ($1 < m < n$) and the same random order as krill like $G_{1,T^c} = \{(W_i, T_i^c), i = 1, \cdots, m\}$ and $G_{2,T^c} = \{(W_i, T_i^c), i = m+1, \cdots, n\}$ and $G_{1,D^c} = \{(W_i, D_i^c), i = 1, \cdots, m\}$ and $G_{2,D^c} = \{(W_i, D_i^c), i = m+1, \cdots, n\}$, respectively. The former groups $G_{1,T^c}$ and $G_{1,D^c}$ are used to detect the optimum threshold to discretize them into binary form, and the latter groups $G_{2,T^c}$ and $G_{2,D^c}$ are discretized into binary using the threshold, and a new contingency table is made for the four types of data (whale, krill, temperature, and depth). Setting the response variable, an expanded procedure for more explanatory variables than the two variables shown in 2.1 is applied to the multi-way contingency table. According to Eqs. (9) and (10), the AICs for all combinations of explanatory variables are given by the conditional probability models. By iterating these procedures 1000 times, the optimum thresholds and AICs can be obtained through averaging their values.

Finally, the best conditional probability model, including the explanatory variables associated with the response variable, is identified. The conceptual outline of the proposed procedure is given by Fig. 1. The numerical procedure including the R package *catdap* (2023) is implemented using R code [R Core Team, 2023] summarized in Supplementary file.

3. Simulation study

To validate our proposed method, the following simulation study was conducted. We first generated 1000 data $x$ by the truncated normal distribution given by

$$f(x;\mu,\sigma,a,b) = \frac{1}{\sigma} \frac{\Phi\left(\frac{x-\mu}{\sigma}\right)}{\Phi\left(\frac{b-\mu}{\sigma}\right) - \Phi\left(\frac{a-\mu}{\sigma}\right)},$$

where $\mu$ and $\sigma$ are mean and standard deviation of the distribution for random variable $X$ within $-\infty \leq a < b \leq \infty$. In this case, the data, called 'simb,' are mixed by two truncated normal distributions $f(x;3,1.75,1,10)$ and $f(x;7,0.75,1,10)$ as case1, $f(x;3,0.75,1,10)$ and $f(x;7,1.75,1,10)$ as case2, and $f(x;3,0.75,1,10)$ and $f(x;7,0.75,1,10)$ as case3. The histogram for each case presents different two peaks summarized in Fig. 2. Then, we generated categorical data, called 'csimb,' which have the same array size as simb:

$$\text{Case 1: csimb} = \begin{cases} 1 & \text{for } 2.0 \leq \text{simb} < 4.0 \text{ or } 6.0 \leq \text{simb} < 8.0 \\ 0 & \text{otherwise} \end{cases},$$

$$\text{Case 2: csimb} = \begin{cases} 1 & \text{for } 2.0 \leq \text{simb} < 4.0 \text{ or } 6.5 \leq \text{simb} < 8.5 \\ 0 & \text{otherwise} \end{cases}, \text{ and}$$

$$\text{Case 1: csimb} = \begin{cases} 1 & \text{for } 2.5 \leq \text{simb} < 3.5 \text{ or } 6.5 \leq \text{simb} < 7.5 \\ 0 & \text{otherwise} \end{cases}.$$

Furthermore, we prepared another type of data, called 'simc,' that includes 1000 data $y$ generated by beta distribution, given by

$$f(y;a,b) = \frac{\Gamma(a+b)}{\Gamma(a)\Gamma(b)} y^{a-1}(1-y)^{b-1},$$

where $\Gamma$ denotes a gamma function, and we set $a = 3.45$ and $b = 10$ in this study.

Summarizing these procedures, we now have the following artificial datasets:

    simb (cases 1, 2, and 3): 1000 real-number data generated by two truncated normal distributions,

    csimb: 1000 binary data by categorization of simb based on two thresholds, and

    simc: 1000 real-number data generated by beta distribution.

We illustrated the histograms for simb and simc to consider the ranges to find the optimum thresholds for two variables as seen in Fig. 2. We set the range to detect the threshold of simb (case1) as 1.5 to 8.0, simb(case2) as 2.0 to 9.0, and simb (case3) as 2.0 to 8.0, and the threshold of simc as 0.1 to 0.5.

The proposed method introduced in 2.3 is applied to the two combinations of csimb and simb and csimb and simc using *catdap* (without histogram model). The csimb is set as the response variable and the simb in case 1, 2, or 3 and simc are set as the explanatory variables. The models are given by

$$\text{Model 1: Model}(\text{csimb; simb(case*), simc})$$
$$\text{Model 2: Model}(\text{csimb; simb(case*)}), \text{ and}$$
$$\text{Model 3: Model}(\text{csimb; simc}).$$

For the outputs, the mean of the optimum thresholds for simb and simc were

    5.96 and 0.28 for case 1,

    4.00 and 0.20 for case2, and

    4.73 and 0.26 for case3,

respectively. Furthermore, the averaged AIC values for the above three models are summarized in Table 1. The averaged AIC for model 2 always becomes the minimum, which is the correct association. While the values between model 1 and model 2 are similar, there is no association from simc by the positive mean value of AIC for model 3. In *catdap*, when AIC indicates zero, the model includes no variable as the explanatory variable. Therefore, if the AIC of a model indicates a positive value, the explanatory variable assumed in the model is basically not dependent on the response variable and is not necessary as an explanatory variable in the model.

This simulation procedure is performed in R (R Core Team, 2023), and simb is generated using the function '*rtruncnorm*' in R (Mersmann et al., 2018) as seen in Supplementary file.

4. Real-data analysis

4.1 Field observations

We applied our methods to two cases. The first uses visual sightings of fin whales from line transect surveys in the Southern Ocean, while the second case uses visual sightings of minke whales from line transect surveys in the in the Arctic Ocean to the west and north of Svalbard.

*Case 1: Southern Ocean fin whales*

Visual observations were carried out onboard three of the six vessels participating in the 2019 krill survey: the *R/V Kronprins Haakon* (KPH), the *F/V Cabo de Hornos* (CDH), and the *RRS Discovery* (DIS). Surveys covered the periods 10 Jan – 22 Feb (KPH), 09 Jan – 11 Mar (CDH) and 03 Jan – 11 Feb (DIS). Observations were taken by dedicated observers, in sea states below Force 6, and generally covered all daylight hours (continuously modified to follow local time and daylength depending on latitude and longitude). On the CDH, only one dedicated observer was present onboard, while two and four dedicated observers were present on the KPH and DIS, respectively. This resulted in a reduced effort especially on the CDH 3 compared to the DIS. To alleviate the problem of under-staffing, dedicated observations on CDH and KPH were generally limited to one forward quadrant (port side, or 270º–360º degrees on the KPH, starboard side, or 0º–90ºon the CDH, relative to the bow of the vessel), with observations in other quadrants recorded as "incidental sightings." On the CDH and KPH, observations were carried out from inside the bridge or an inside observation deck, and sightings were taken as voice recordings directly to disc, using the system developed for the Norwegian surveys for North Atlantic minke whales (Øien, 1995). This system allows the dedicated observer to record effort, weather, and sightings through a handheld microphone, while maintaining full visual attention. Observations on the DIS were carried out from outside platforms by two dedicated observers and one data recorder. Data were entered into the Logger software system (http://www.marineconservationresearch.co.uk/downloads/logger-2000-rainbowclick-software-downloads/). Standard variables were recorded, including estimated radial distance, angle relative to the vessel's heading, species, group size, swimming direction, and initial cue. Radial distance was estimated using either 7 x 50 reticulated binoculars or (on the KPH) 30-cm equidistant steps on a mast ladder positioned 16.6 meters forward of the observation deck. These steps correspond to different angles of depression relative to the horizon, calibrated for the height of each observer. Essentially, this method follows the exact same logic as that of reticulated binoculars or a distance stick. Angle relative to the bow was determined using a standard angle board. Weather and sea-state were recorded every 15–30 minutes, and in some cases (KPH) detailed weather station data were available from the ship's automatic data recording system. Transects were split into 1 nm long segments, and the number of fin whale sightings (groups and individuals) were summarized within each segment. All positive sightings (i.e. number of individuals within a segment greater than or equal to 1). We also included surface temperature (Celsius) and water depth (meter) for a 1-nm segment to investigate the association from environmental data for the biological community. Figure 3a gives spatial plots for krill biomass (krill), sighting data of fin whales (w), surface temperature (sst), depth data (depth), slope for the depth (slope), and gradient surface temperature (sstgrd).

Data on water depth came from the ETOPO 1 bathymetric dataset available at https://www.ncei.noaa.gov/products/etopo-global-relief-model. Data were extracted for the middle position of each 1-nm segment throughout the survey tracks. Data on sea surface temperature sst were obtained from the OISST dataset available at https://www.ncei.noaa.gov/products/optimum-interpolation-sst and extracted in the same way as for depth data. The slope for the depth is the maximum rate of change in the depth from that cell to its neighbours calculated by the *Slope* tool of the Surface toolset in ArcGIS https://desktop.arcgis.com/en/arcmap/10.3/tools/spatial-analyst-toolbox/an-overview-of-the-surface-tools.htm. The lower the slope value, the flatter the terrain; the higher the slope value, the steeper the terrain. The gradient surface temperature is also calculated by *Slope* applied to sst.

*Case 2: Arctic Ocean minke whales*

During the years 2014–2017, ecosystem surveys were conducted in August-September in the Arctic Ocean to the west and north of Svalbard (Solvang et al., 2021). Sampling included all trophic levels from phytoplankton to whales, as well as chemical and physical properties of the water masses in the area. In Solvang et al. (2021), the associations among minke whale, the predatory fish Atlantic cod, and 0-group fishes and zooplankton as their potential prey species were investigated. The data of minke whale are the counted number and the data of Atlantic cod, 0-group fishes, and zooplankton are acoustic registration as the nautical area scattering strength [$S_A$, dB re 1 m$^2$nmi$^{-2}$, $S_A = 10\log 10(s_A)$]. Figure 3b presents the data for a grid cell size of 50 km according to the transects of cruise and the aggregated upper 200 m in the depth.

4.2. Preliminary analyses

Before applying the proposed procedure to the data, the relationships among variables are first investigated by regression models. The counted number of sighted whales is converted into binary data, 1 for number > 0 and 0 for number = 0. The logistic regression model for whales and the linear regression model for the continuous response variable are considered. The best model is selected by minimum AIC. The effect size is also investigated. The logistic and linear regression analyses are conducted by *glm* and *lm* in R (R Core Team, 2023).

*Case 1.*

Krill biomass showed a high distribution close to zero because there are many locations indicated by zero as shown in Fig. 3, *Case1*. Thus, logarithmic transformation of krill observations is required for the data by replacing zero with a small real number. We consider the logistic regression model (model w) for the binary data of whale and the linear regression model (model k) for the logarithmic-transformed krill biomass data. The best models selected by minimum AIC are

Model w: Appearance probability of fin whale = $(1 + \exp(-0.76 + 0.014 \times \log(krill) - 0.21 \times sst - 3.4\text{e-}05 \times Depth + 0.058 \times slope))^{-1}$, and

Model k: $\log(krill) = -7.1 + 1.2 \times$ as factor (binary data of fin whale) $+ 4.2\text{e-}04 \times Depth - 0.41 \times SST + 0.26 \times slope$.

The p-values for the estimated coefficients in model w are 0.00050 for log(krill), 1.06e-09 for SST, 0.12 for Depth, and 3.01e-06 for slope, that is, three explanatory variables (except for Depth) were significantly associated with the appearance of fin whale. On the other hand, the p-values for the estimated coefficients in model k are 0.00056 for whale, 4.8e-06 for depth, 0.0020 for sst, and 1.8e-06 for slope, that is, all explanatory variables were significantly associated with log(krill).

In multinomial logistic regression (model w), the impact of predictor variables in a logistic regression model is explained in terms of the odds ratio, which reflects the effect size measures (Aziz et al., 2016; Field, 2009). The effect sizes for log(krill), SST, Depth, and Slope are given by the odds below:

log(krill): $\exp(0.014) = 1.02$,

sst: $\exp(-0.12) = 0.79$,

depth: $\exp(3.4\text{-e}05) = 1$, and

slope: $\exp(0.058) = 1.06$.

The effect sizes for the appearance of fin whales indicate a low value, 2% and 6% effects for increases in 1.0 logarithmic krill biomass and 1.0 meter for slope, respectively, and no effect for depth, while the effect size of appearance of fin whales shows 21% negative effects for an increase of 1° Celsius.

For the output by linear model, the eta squared, which is calculated by the effect size of ANOVA (Ben-Shachar et al., 2020) indicates small effect sizes for all explanatory variables as below:

binary data of counted number of fin whales: 4.9e-03,

depth:      6.0e-03,

sst:          4.2e-03, and

slope:       6.0e-03.

*Case 2.*

Solvang et al. (2021) described the logistic regression analysis used to investigate whether certain prey groups were more or less likely to be present or absent when minke whales were present. The modelling was conducted for each prey species. In comparing the models including two and three explanatory variables, the minimum AIC selected the model that includes only cod given by

Appearance probability of minke whale $= (1 + \exp(-2.4 + 0.0028 \times \text{cod}))^{-1}$.

The effect size is $\exp(0.0028) = 1.0028$, which is a low value.

4.3. Results by proposed procedure

First, we describe histograms of the observations in Fig. 4 to find the region to use in searching for the optimum threshold to obtain two-category data. In Case 1, Krill biomass showed a high distribution close to zero because there are many locations indicated by zero. Thus, logarithmic transformation of krill observations is required for the data by replacing zero with a small real number. The histogram of logarithmic-transformed data showed a dense distribution around small values and a long-tailed distribution with a single peak. The range for finding the optimum threshold for making two-category data is set from -3 to 8 (i.e., $a = -3$ and $b = 8$ in $[a,b]$ of subsection 2.3). For the surface temperature (sst), the histogram seems to have two different distributions around 2.0° Celsius. The optimum threshold to make two categorical data is searched for within the range $[-2.0, 2.5]$ (degrees Celsius). For the water depth (depth), we set the range to $[-1,000, 0.0]$ (meter), where negative means under the sea surface. The slope (slope) and temperature gradient (sstgrd) are set to $[1,15]$ (meter) and $[0.00001, 0.0017]$ (degrees Celsius), respectively. In Case 2, we set the range to $[2.0, 9.9]$ for logarithmic zooplankton biomass (plk), $[-7.5, 9.0]$ for logarithmic cod biomass (cd), and $[0.1, 9.9]$ for logarithmic 0-group fishes biomass (0gr).

In applying our proposed method to these data, the averaged optimum thresholds are 2.39 for log(krill), 0.99° Celsius for sst, -386.1 meters for depth, 6.52 meters for slope, and 0.00087° Celsius for sstgrd in Case 1, and 4.41 for log(plk), 3.94 for log(cd), and 2.79 for log(0gr) in Case 2. Using the optimum thresholds, we could obtain the categorical data for log(krill), sst, depth, slope, and sstgrd in Case 1 and for log(plk), log(cd), and log(x0g) in Case 2. Those data and the categorical data for whale are integrated, and the R function *catdap* is applied to the dataset. The response variable is set as whale or log(krill) in Case 1, and it is set as whale, plk, cd or x0g in Case 2. Against the response variables, we use several combinations of explanatory variables, e.g., krill, sst, and depth for whale (in Case 1). Tables 2 and 3 present the calculated AICs for all possible models in Case 1 and Case 2, respectively. Using these categorical data, the minimum AIC selects the model including sst, depth, and slope for fin whale, and the model including depth, slope, and sstgrd for krill in Case 1 (Table 2). In Case 2, the minimum AIC selects the model including plk and cd for minke whale, the model including whale and cd for pl, the model including pl and 0gr for cd, and the model including pl and cd for 0gr (Table 3a). Table 3b summarizes the calculated AICs for all possible models among the prey species of minke whale. The minimum AICs indicate that the species are associated with each other. Based on the minimum AIC models for whale and krill, we diagram the associations among the four types of data in Fig. 5.

Furthermore, we demonstrate how to apply the histogram model, which has been set for continuous variables in *catdap*. When setting fin whale as the response variable in Case 1, it automatically identifies 14 categories for sst, 26 categories for depth, 5 categories for slope; meanwhile, 3 categories are identified for krill and 1 category for sstgrd. Using these categorical data given by *catdap*, the minimum AIC selects the explanatory model including sst, depth and slope when the response variable is krill. For whale as a response variable, the minimum AIC selects the model including sst and depth. When setting minke whale as the response variable in Case 2, *catdap* automatically identifies 3 categories for plk, cd, and 0gr. The minimum AIC selects the model including plk and cd, as done in our method.

5. Discussion

In our method, the optimum threshold to divide continuous data into two types of categorical data is found by taking account of the associations with the response variable, e.g., whale or krill. The obtained threshold for the logarithmic krill biomass is 2.39, which is among the highest frequencies of the histogram in Fig. 4, Case 1. A value less/larger than 2.39 is interpreted as a lower/higher value for the logarithmic biomass. The optimum threshold for surface temperature is 1.01, which is interpreted as a temperature warmer than 1.0° Celsius and a lower temperature of less than 1.0° Celsius. The optimum depth threshold is -386.1, which seems reasonable given the common knowledge of the fin whale's diving depth (Fronseca et al. 2022). The optimum threshold for the logarithmic plankton biomass is 4.41, which is among the highest frequencies of the histogram in Fig. 4, Case 2. On the other hand, the optimum thresholds 3.94 for the logarithmic cod biomass and 2.79 for the logarithmic 0-group fish biomass seem to divide the values taking higher frequencies in the histograms.

For the output by *catdap* presented in Table 2, the associations of the slope with whale and krill may be caused by the fact that continental slopes often create frontal systems that gather biological material, thereby attracting krill and other planktonic organisms. While the preliminary regression analysis showed a significant effect from logarithmic krill biomass for the appearance of fin whale, the relationship between fin whale and krill may be indirectly associated through the association with the slope. For the output by *catdap* presented in Table 3, the association for whale with cod may be related to the association from zooplankton to whale/cod. The preliminary regression analysis showed the significant effect from cod to the appearance of common minke whale. The output by our approach would support the suggestion that there exists a connection between cod abundance and feeding conditions for other top predators such as common minke whales (Solvang et al., 2022).

While the models for the secondary minimum AICs in Case 1 ( P(whale | sst, depth, slope): 2177.1, P(whale | sst, slope, sstgrd): 2177.0) do not include whale or krill, the models for the minimum AICs in Case 2 show the association for minke whale with plankton (zooplankton). The diagram in Case 2 supports the interpretation that plankton, cod, and 0-group fish are influenced by minke whales (Solvang et al., 2021). Furthermore, the secondary minimum AIC (99.4) of the model P(plk | cd, 0gr) is very close to the minimum AIC (98.9) of the model P(plk | 0gr), which the diagram presents the association for cod with plankton in the right-hand side of upper figures in Fig. 5, Case 2. The diagram for the association among only prey species shows the lower figure in Fig 5. The secondary minimum AICs models to plankton and cod are very close to the minimum AICs and the output summarizes that the three prey species associated with each other.

Increasing demand for commercial harvesting of krill requires a careful assessment of sustainable harvest levels. As krill-dependent predators, several Southern Hemisphere populations of humpback whales (*Megaptera novaeangliae*) have undergone dramatic population recoveries in recent decades (Baines et al., 2021). Fin whales in the Southern Hemisphere were the most heavily exploited in terms of numbers taken during the period of intense industrial whaling, but the recent abundance of fin whales also suggests that they are undergoing a substantial recovery (Herr et al., 2022). The results of our analysis reflect the recent tendency between fin whales and krill biomass, and they suggest that our proposed method could contribute to the CCAMLR risk assessment and future management systems for Antarctic krill.

6. Conclusion

We presented a categorical data analysis by combining the procedure (CATDAP) provided by Sakamoto and Akaike (1978), with a method to replace continuous values by two types of categorized data. The optimum threshold obtained by training data is used to categorize continuous explanatory variables to binary data using test data. This procedure helps to avoid complicated categorization by a histogram model as conducted for continuous variables in the calculation of *catdap*. The proposed procedure is iterated a large enough number of times, and the averaged threshold is used as the optimum threshold. The averaged AIC is also used for finally determining whether the response variable is dependent on the explanatory variable between the paired data. Using the spatial temporal data obtained by two scientific surveys, our proposed method attempts to objectively investigate the likelihood of several events, such as the number of overlapped locations among data across months or years. Knowing the associations among variables estimated by this method would be useful before applying a more specific model to estimate causality such as a directed acyclic graph (Pearl, 1998, 2009), which has been widely applied to Bayesian networks (Koller and Friedman 2009, Neapolitan 2003). This method is expected to be used as a practical monitoring tool in integrated ecosystem assessment, especially for following the temporal changes of associations among high-dimensional multivariate data for biological communities and oceanographic data reflecting the impact of human activities.

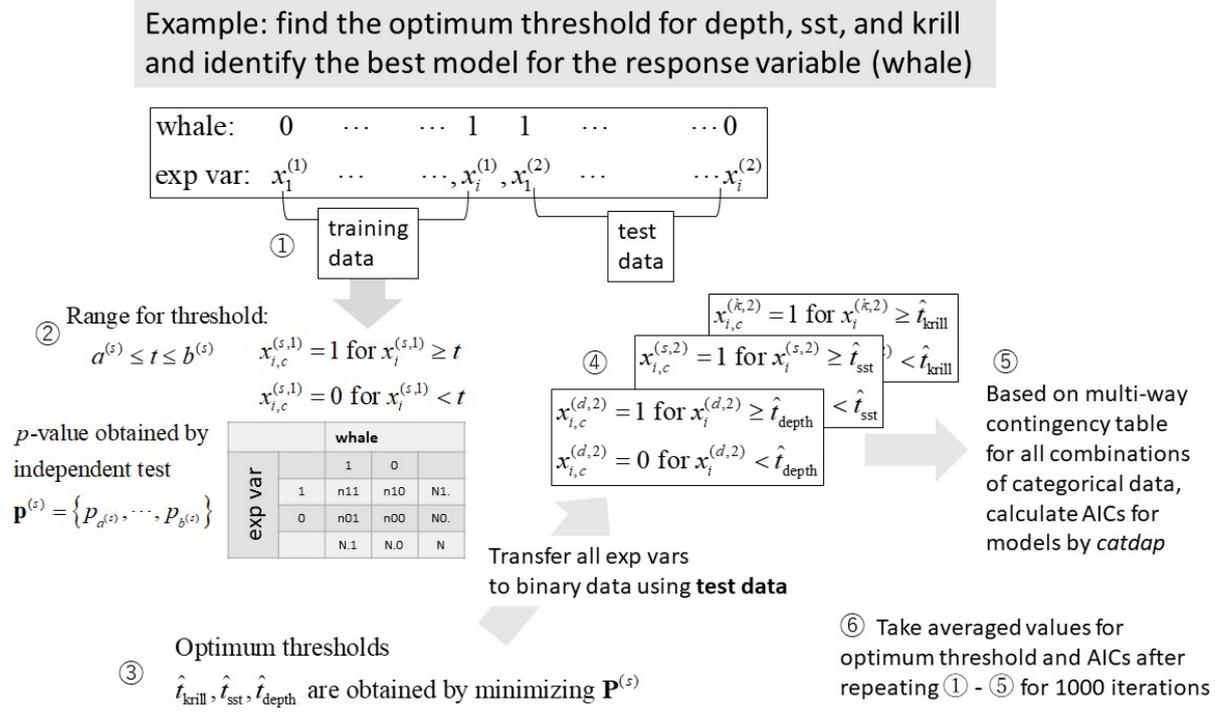

Fig. 1. Flow of proposed procedure. ① Set a continuous explanatory variable for a response variable, e.g., whale, and split it into training set and test set; ② Categorize the explanatory variable for the training data according to a range for the threshold and make a two-way table using the categorized data and the corresponding whale data. Apply an independent test for the table and calculate a p-value for each threshold; ③ Minimize the vector summarizing the obtained p-values and obtain the optimum threshold in the case where the p-value indicates minimum. Step (i) in the text corresponds to ①, ②, and ③. Apply step (i) to all explanatory variables; ④ Transfer all explanatory variables to binary data for the test data using the optimum thresholds; ⑤ Based on a multi-way contingency table for all combinations of the binary explanatory variables, calculate AIC using *catdap*. Step (ii) in the text corresponds to ④ and ⑤. ⑥ Take the averaged values for the optimum threshold and AIC after repeating steps (i) and (ii) 1000 times. The averaged AIC is used to evaluate which model is better among model candidates. The optimum threshold for the better model is used as a final optimum threshold to divide the continuous values into two-category data, 1 and 0. Step (iii) in the text corresponds to ⑥.

Case 1:

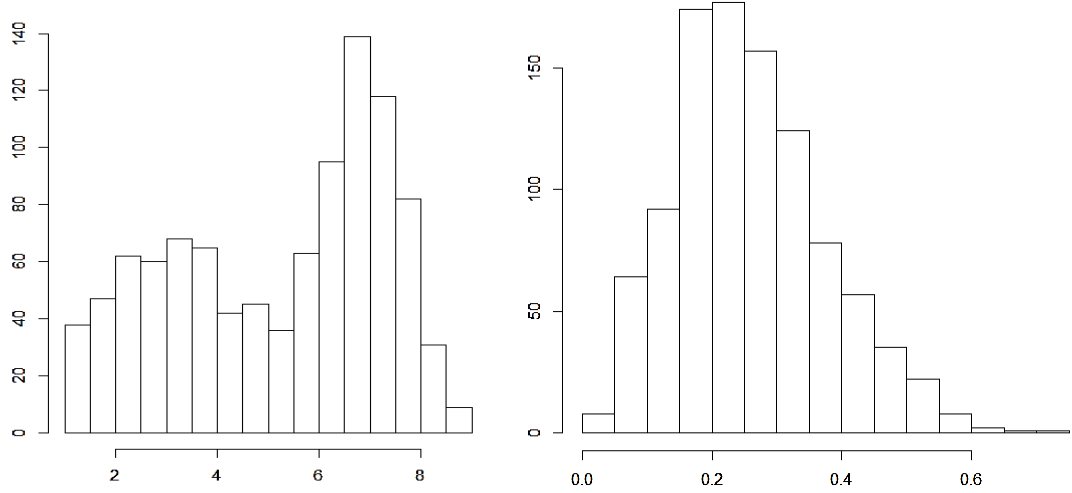

Case 2:

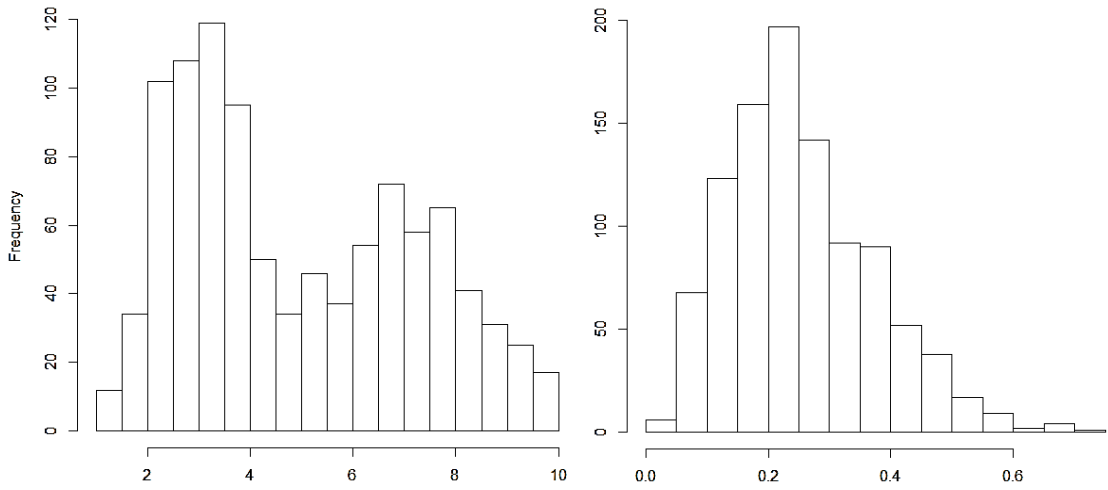

Case 3:

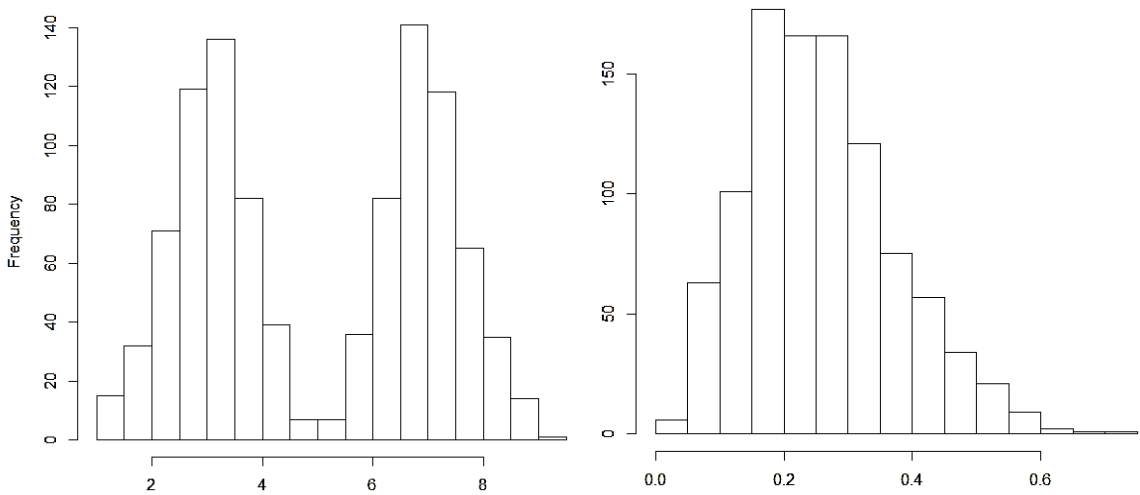

Fig. 2. Histograms of simb generated by two truncated normal distributions (upper) and simc generated by beta distribution (lower).

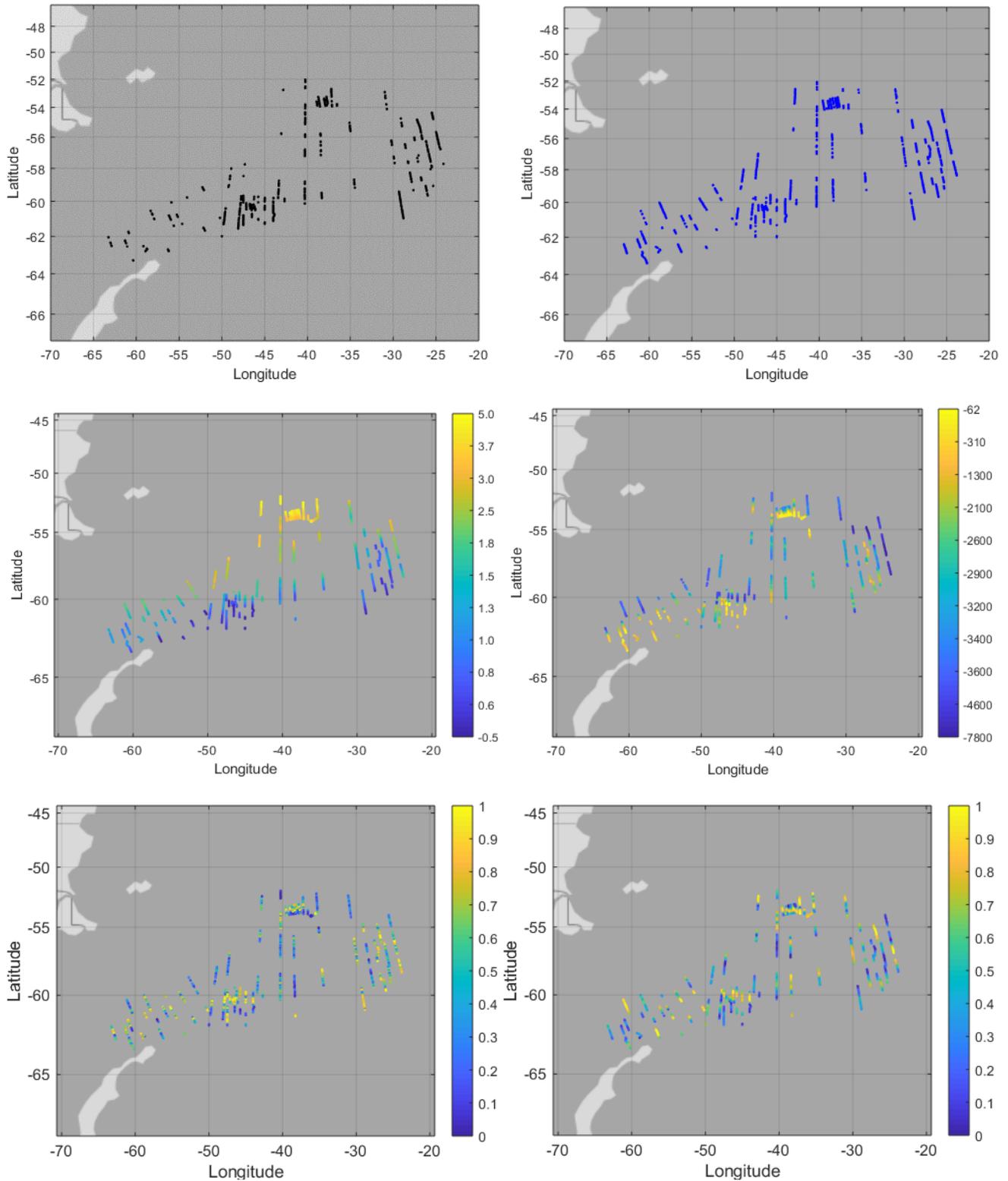

Fig. 3a. Field observations for Case 1. Upper-left: sighting data of fin whales (counted number), upper-right: krill biomass ($gm^{-2}$), middle-left: surface temperature (celsius), middle-right: depth data (meter), lower-left: slope for depth (meter), and lower-right: gradient temperature (celsius). Brighter/darker dots' colors of surface temperature and water depth mean higher/lower temperature and shallower/deeper water depth, respectively.

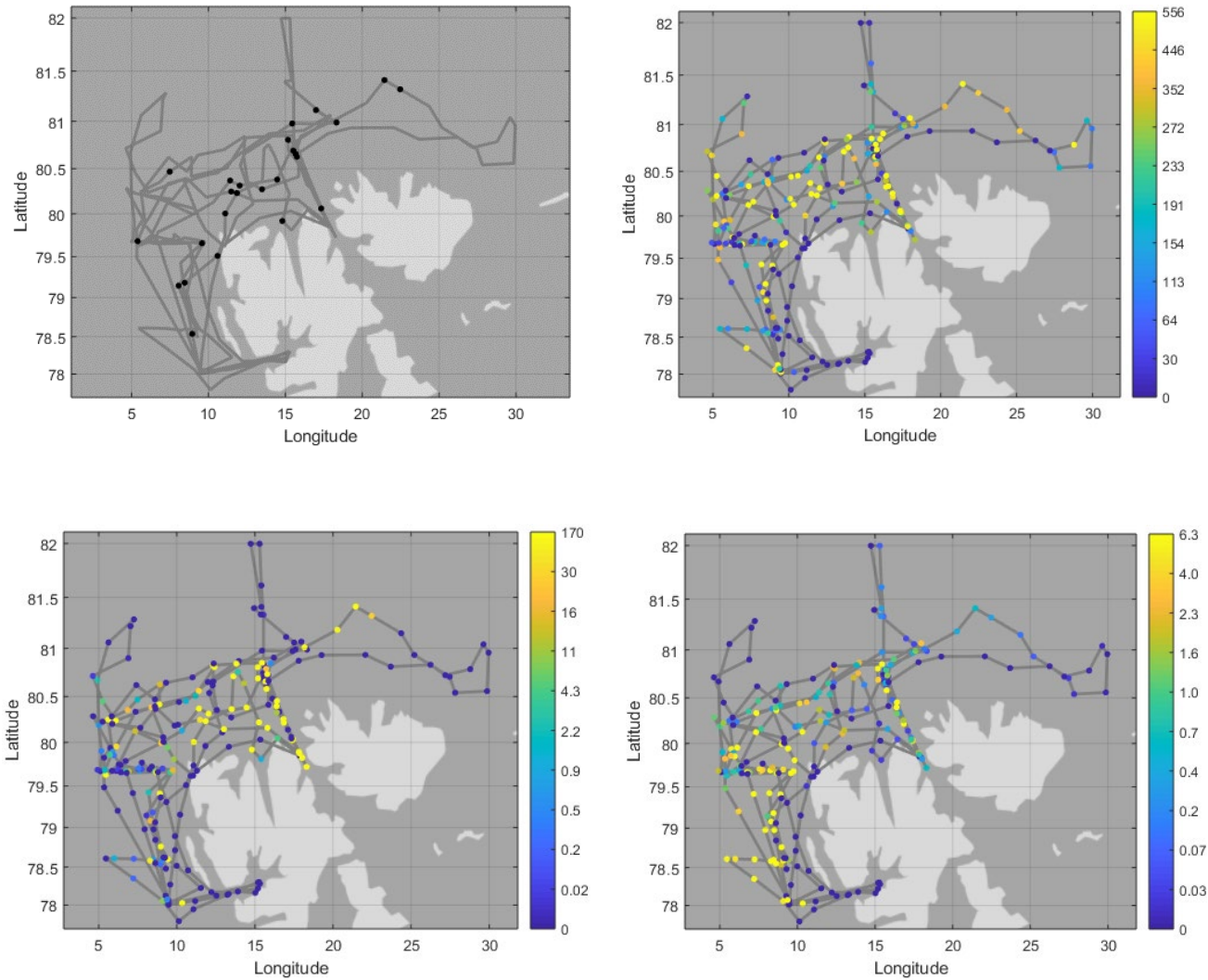

Fig. 3b. Field observations for Case 2 along cruise track during autumn 2014–2017. Upper-left: sighting data of minke whales (counted number), upper-right: integrated values of $s_A$ ($m^2 nmi^{-2}$) for plankton in the upper 200-m depth, lower-left: integrated values of $s_A$ ($m^2 nmi^{-2}$) for Cod in the upper 200-m depth, and lower-right: integrated values of $s_A$ ($m^2 nmi^{-2}$) for 0-group fish in the upper 200-m depth. Brighter/darker dots' colors of higher/lower integrated values of $s_A$.

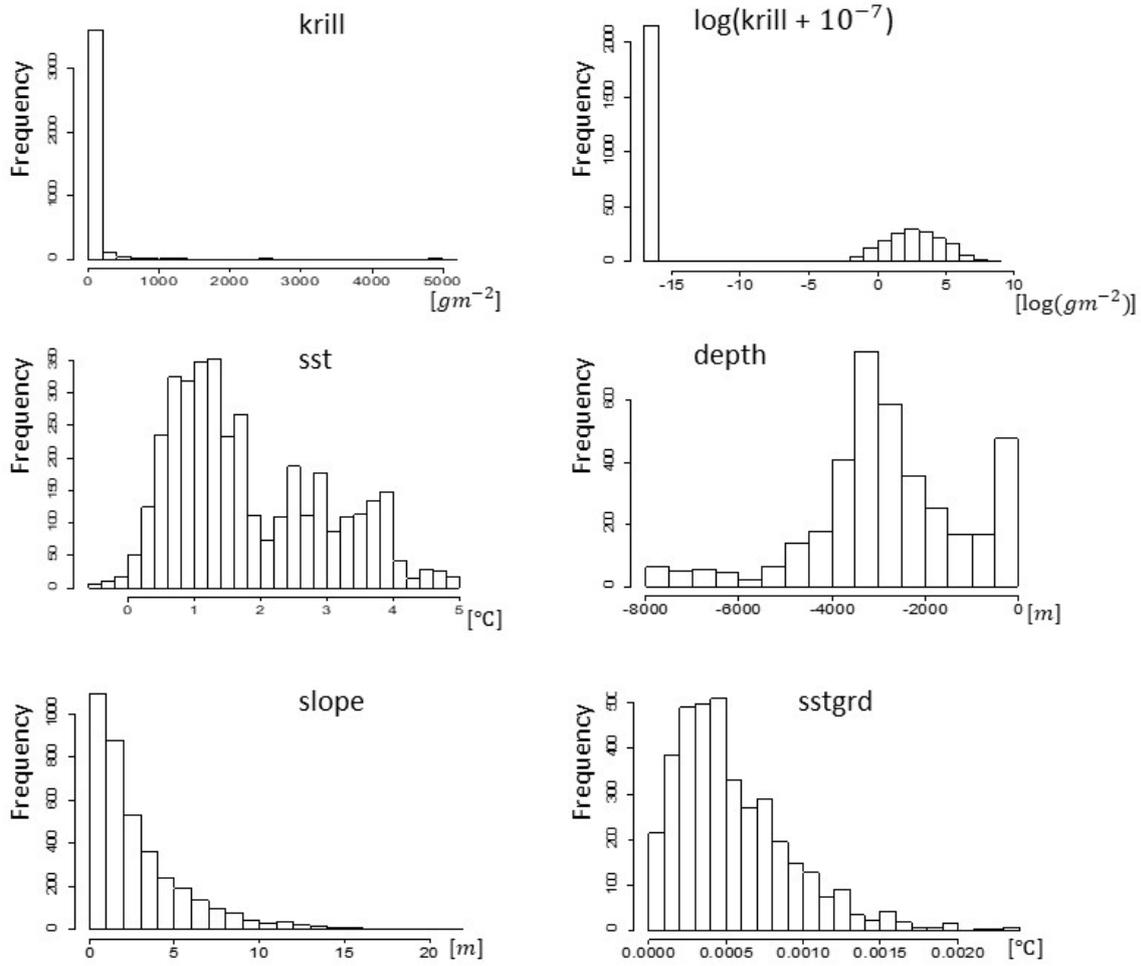

Case 1: Histograms of observations for krill biomass (upper-left, x-axis: biomass $[gm^{-2}]$, y-axis: frequency), logarithmic transformation of krill biomass (upper-right, x-axis: logarithmic-transformed biomass value, y-axis: frequency), surface temperature (middle-left, x-axis: degree Celsius, y-axis: frequency), depth (middle-right, x-axis: meter, y-axis: frequency), slope (lower-left, x-axis: meter, y-axis: frequency), sst gradient (lower-right, x-axis: degree Celsius, y-axis: frequency).

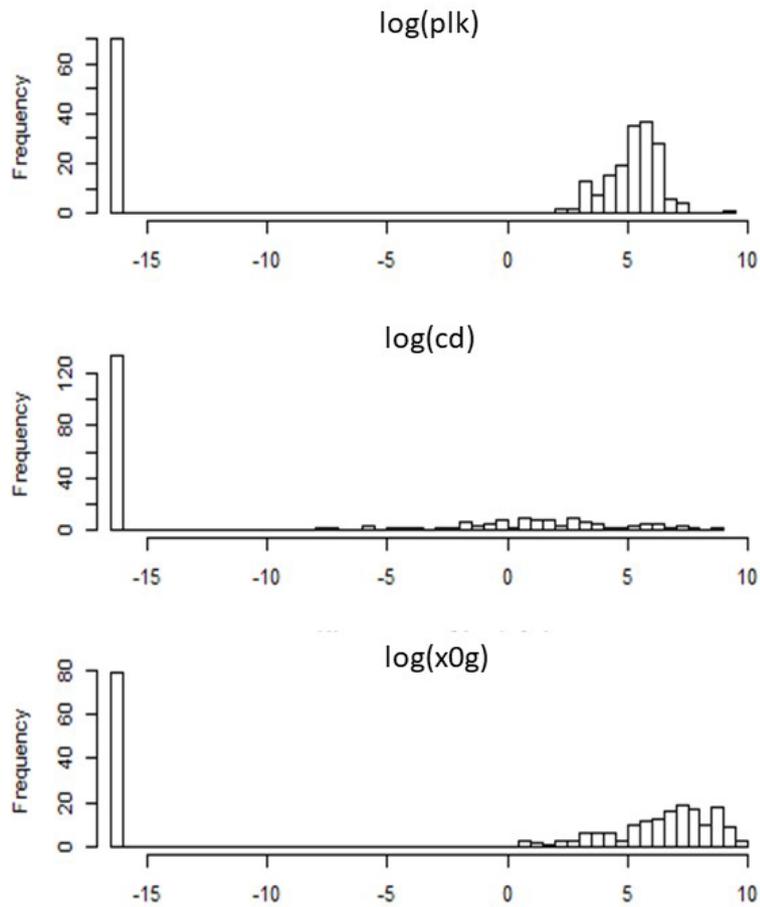

Case 2: Histograms of observations for logarithmic transformation of plankton biomass (top, x-axis: logarithmic-transformed biomass [g m-2], y-axis: frequency), logarithmic transformation of cod biomass (middle, x-axis: logarithmic-transformed biomass value, y-axis: frequency), and logarithmic transformed 0-group fish (bottom, x-axis: meter, y-axis: frequency).

Fig. 4. Histograms of observations for Cases 1 and 2. These are used for setting the range $[a,b]$ to detect the optimum threshold in a computational procedure.

Case 1

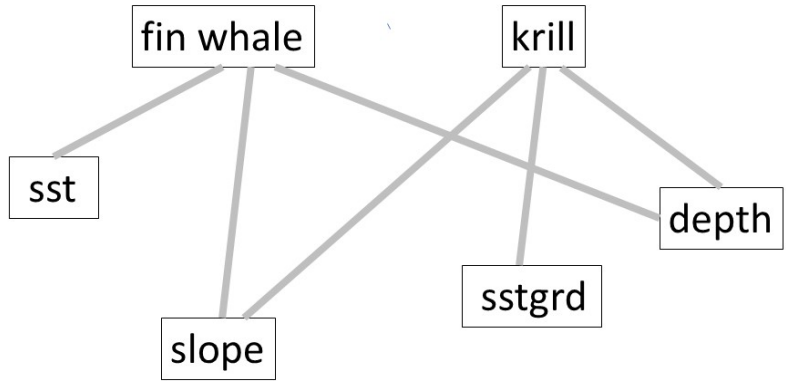

Case 2

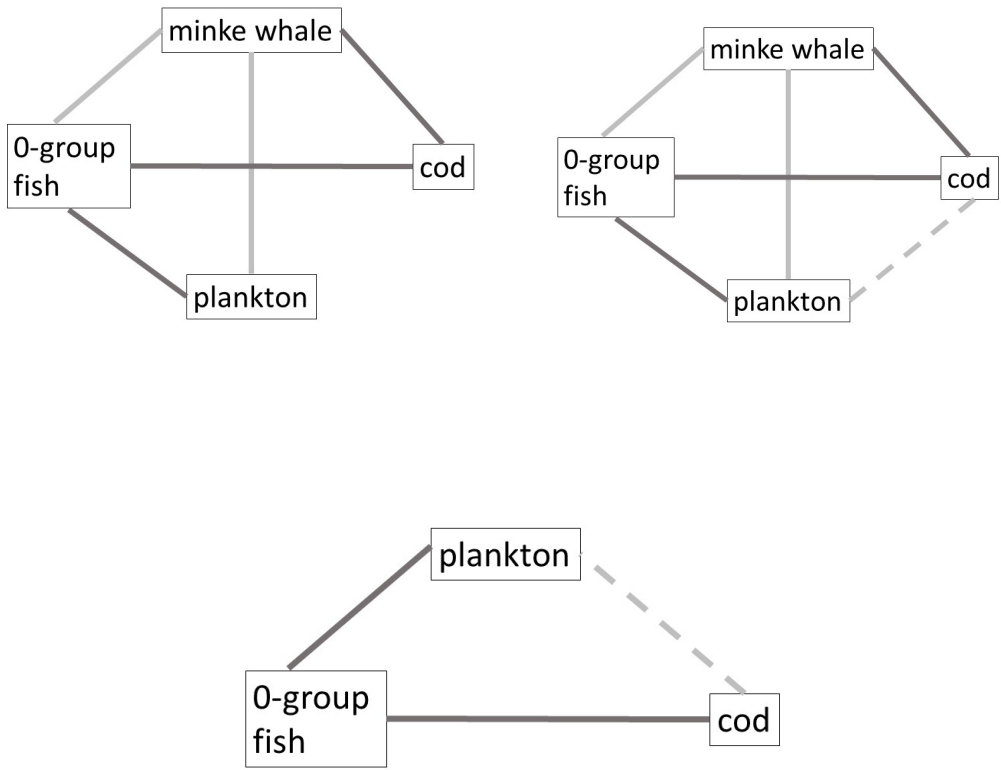

Fig. 5. Diagrams of the associations based on the model indicating minimum AIC by *catdap*. Case 1: among fin whale, krill, surface temperature, depth, slope, and surface temperature gradient; Case 2: among minke whale, plankton, cod, and 0-group fish. Left diagram is made by the minimum AIC. The upper diagrams are made by the minimum and the secondary minimum AICs (dashed lines). The difference between the minimum AIC and the secondary minimum AIC is less than 2. Dark-coloured lines indicate that both presences are associated with each other. The lower diagram indicates among prey species.

Table 1. Mean values of AIC for the models. Case 1, 2, and 3 correspond to the combinations for explanatory variables, simb in case1 and simc, simb in case2 and simc, and simb in case3 and simc. The response variable of the model is csimc.

|         | Case 1 | Case 2 | Case 3 |
|---------|--------|--------|--------|
| Model 1 | 507.45 | 520.46 | 661.77 |
| Model 2 | 505.16 | 517.93 | 659.62 |
| Model 3 | 622.58 | 641.97 | 695.37 |

Table 2. AIC results obtained by *catdap* applied to the categorical data based on the averaged optimum threshold in Case 1.

For the response variable (whale or krill), several explanatory variables listed in 'From' are considered. For whale, the model explained by sst, depth, and slope is the best. On the other hand, for krill, the model explained by depth, slope, and sstgrd is the best. The relationships are illustrated in Fig. 4.

| | | Response variable | | | | | |
|---|---|---|---|---|---|---|---|
| | | To fin whale | | | | To krill | |
| From | 1var | krill | 2227.4 | From | 1var | whale | 1728.0 |
| | | sst | 2188.5 | | | sst | 1729.9 |
| | | depth | 2227.6 | | | depth | 1716.1 |
| | | slope | 2217.5 | | | slope | 1730.3 |
| | | sstgrd | 2231.5 | | | sstgrd | 1730.3 |
| | 2var | krill, sst | 2184.6 | | 2var | whale, sst | 1726.0 |
| | | krill, depth | 2226.0 | | | whale, depth | 1714.5 |
| | | krill, slope | 2215.3 | | | whale, slope | 1728.1 |
| | | krill, sstgrd | 2228.1 | | | whale, sstgrd | 1726.9 |
| | | sst, depth | 2187.0 | | | sst, depth | 1716.6 |
| | | sst, slope | 2179.8 | | | sst, slope | 1727.3 |
| | | sst, sstgrd | 2185.1 | | | sst, sstgrd | 1728.1 |
| | | depth, slope | 2213.1 | | | depth, slope | 1714.0 |
| | | depth, sstgrd | 2228.2 | | | depth, sstgrd | 1715.8 |
| | | slope, sstgrd | 2219.0 | | | slope, sstgrd | 1730.2 |
| | 3var | krill, sst, depth | 2186.4 | | 3var | whale, sst, depth | 1715.9 |
| | | krill, sst, slope | 2178.5 | | | whale, sst, slope | 1726.1 |
| | | krill, sst, sstgrd | 2183.2 | | | whale, sst, sstgrd | 1725.4 |
| | | krill, depth, slope | 2212.6 | | | whale, depth, slope | 1713.6 |
| | | krill, depth, sstgrd | 2227.4 | | | whale, depth, sstgrd | 1714.8 |
| | | krill, slope, sstgrd | 2217.4 | | | whale, slope, sstgrd | 1728.8 |
| | | sst, depth, slope | 2177.1 | | | **sst, depth, slope** | **1712.7** |
| | | sst, depth, sstgrd | 2183.3 | | | sst, depth, sstgrd | 1715.3 |
| | | sst, slope, sstgrd | 2177.0 | | | sst, slope, sstgrd | 1726.0 |
| | | depth, slope, sstgrd | 2214.4 | | | depth, slope, sstgrd | 1714.9 |
| | 4var | krill, sst, depth, slope | 2180.2 | | 4var | whale, sst, depth, slope | 1715.8 |
| | | krill, sst, depth, sstgrd | 2186.1 | | | whale, sst, depth, sstgrd | 1717.4 |
| | | krill, sst, slope, sstgrd | 2179.0 | | | whale, sst, slope, sstgrd | 1727.6 |
| | | krill, depth, slope, sstgrd | 2215.6 | | | whale, depth, slope, sstgrd | 1716.1 |
| | | **sst, depth, slope, sstgrd** | **2174.9** | | | sst, depth, slope, sstgrd | 1712.9 |
| | 5var | krill, sst, depth, slope, sstgrd | 2182.3 | | 5var | whale, sst, depth, slope, sstgrd | 1728.4 |

Table 3. AIC results obtained by *catdap* applied to the categorical data based on the averaged optimum threshold in Case 2.

a. For the response variable (minke whale, plankton, cod, and 0gr), several explanatory variables listed in 'From' are considered. For minke whale, the model explained by the presence of plankton and cod is the best. For plankton, the model explained by the presence of cod and 0gr is the best. For cod, the model explained by the presence of plankton and 0gr is the best. Finally, for 0gr, the model explained by the presence of plankton and cod is the best. The relationships are illustrated in Fig. 4.

|  |  | Response variable | | | | | | | |
|---|---|---|---|---|---|---|---|---|---|
|  |  | To minke whale | | To plankton | | To cod | | To 0gr | |
| Explanatory variable | 1 | PL | 81.0 | W | 143.1 | W | 66.2 | W | 155.4 |
|  |  | CD | 78.5 | CD | 138.2 | PL | 66.8 | PL | 114.0 |
|  |  | 0gr | 78.2 | **0gr** | **98.9** | 0gr | 62.4 | CD | 152.0 |
|  | 2 | PL, CD | 80.0 | W, CD | 140.9 | W, PL | 63.6 | W, PL | 115.3 |
|  |  | PL, 0gr | 81.5 | W, 0gr | 102.1 | **W, 0gr** | **61.5** | W, CD | 151.6 |
|  |  | **CD, 0gr** | **77.4** | CD, 0gr | 99.4 | PL, 0gr | 63.0 | **PL, CD** | **112.7** |
|  | 3 | PL, CD, 0gr | 81.5 | W, CD, 0gr | 105.0 | W, PL, 0gr | 63.9 | W, PL, CD | 116.3 |

b. For the response variable (plankton, cod, or 0gr) in prey species, several explanatory variables listed in 'From' are considered. The best model for each response variable is explained by the presence of the other two species.

|  |  | Response variable | | | | | |
|---|---|---|---|---|---|---|---|
|  |  | To plankton | | To cod | | To 0gr | |
| Explanatory variable | 1 | CD | 138.2 | PL | 63.8 | PL | 114.0 |
|  |  | **0gr** | **98.9** | **0gr** | **62.4** | CD | 152.0 |
|  | 2 | CD, 0gr | 99.4 | PL, 0gr | 63.0 | **PL, CD** | **112.7** |

References

Akaike, H. 1974. A new look at the statistical model identification. IEEE Transactions on Automatic Control, 19: 716-723.

Baines, M., Kelly, N., Reichelt, M., Lacey, C., Pinder, S., Fielding, S., Murphy, E., Trathan, P., Biuw, M., Lindstrøm, U., Krafft, B., and Jackson, J. 2021. Population abundance of recovering humpback whales *Megaptera novaeangliae* and other baleen whales in the Scotia Arc, South Atlantic. Marine Ecology Progress Series, 676; 77-94. https://doi.org/10.3354/meps13849.

Ben-Shachar M.S, Lüdecke D., Makowski D. 2020. effectsize: Estimation of Effect Size Indices and Standardized Parameters. Journal of Open Source Software, **5**(56), 2815. doi:10.21105/joss.02815, https://doi.org/10.21105/joss.02815.

Field, A. 2009. Discovering Statistics Using SPSS. Third Edition. SAGE Publications Ltd.

Fonseca, C.T., Pérez-Jorge, S., Prieto, R., Oliveria, C., Tobeña, M., Scheffer, A., and Silva, M.A. 2022 Dive behavior and activity patterns of fin whales in a migratory habitat. Frontiers in Marine Science, https://doi.org/10.3389/fmars.2022.875731.

Herr, H., Viquerat, S., Devas, F., Lees, A., Wells, L., Gregory, B., Giffords, T., Beecham, D., and Meyer, B. 2022. Return of large fin whale feeding aggregations to historical whaling grounds in the Southern Ocean. Scientific reports. 12; 9458. https://doi.org/10.1038/s41598-022-13798-7.

Katsura, K. and Sakamoto, Y. 1980. Computer Science Monograph, No.14, CATDAP, A Categorical Data Analysis Program Package. The Institute of Statistical Mathematics.

Koller, D. and Friedman, N. 2009. Probabilistic Graphical Models: Principles and Techniques. MIT Press.

Krafft, B., Bakkeplass, K., Berge, T., Biuw, M., Erices, J., Jones, E., Knutsen, T., Kubilius, R., Kvalsund, M., Lindstrøm, U., Machaulay, G.J., Renner, A., Rey, A., Skern, R., Søiland, H., Wienerroither, R., Ahonen, H., Goto, J., Home, N., Huerta, M., Höferm J., Iden, O., Jouanneau, W., Kruger, L., Håvard, L., Lowther, A., Makhado, A., Mestre, M., Narvestad, A., Oosthuisen, C., Rodringues, J., and Øyerhamn, R. 2019. Report from a krill focused survey with RV Kronprins Haakon and land-based predator work in Antarctica during 2018/2019. Rapport fra Havforskningen. 21, ISSN:1893-4536.

Krafft, B.A., Macaulay, G.J., Skaret, G. Knutsen, T., Bergstad, O.A., Lowther, A., Huse, G., Fielding, S., Trathan, P. Murphy, E., Choi, SG., Chung, S., Han, I., Lee, K., Zhao, X., Wang, X., Ying, Y., Yu, X., Demianenko, K., Podhornyi, V., Vishnyakova, K., Pshenichnov, L., Chuklin, A., Shyshman, H., Cox, M.J., Reid, K., Watters, G.M., Reiss, C.S., Hinke, J.T., Arata, J., Godø, O.R., and Hoem, N. 2021. Standing stock of Antarctic krill (*Euphausia superba* Dana, 1850) (Euphausiacea) in the Southwest Atlantic sector of the Southern Ocean, 2018-19. Journal of Crustacean Biology. 41(3); 1-17. https://doi.org/10.1093/jcbiol/rnab046.

Lakens, D. 2013. Calculating and reporting effect sizes to facilitate cumulative science: a practical primer for t-tests and ANOVAs. Review Article, 4;Article 863. https://doi.org/10.3389/fpsyg.2013.00863.

Macaulay, G., Skaret, G., Knutsen, T., Bergstad, O.A., Krafft, B., Fielding, S., Choi, S., Chung, S., Demianenko, K., Podhornyi, V., Vishnyakova, L., Pshenichnov, A., Chuklin, A., Shishman A., Wang, X., Zhao, X., and Cox, M. 2019. Biomass results from the International Synoptic Krill Survey in Area 48. SG-ASAM-2019, 08.

Mersmann, O., Trautmann, H., Steuer, D. and Bornkamp, B. 2018. Package 'truncnorm', CRAN.

Neapolitan, R.E. 2003. Learning Bayesian Networks. Prentice Hall.

R Core Team. 2023. R: A language and environment for statistical computing. R Foundation for Statistical Computing, Vienna, Austria. URL https://www.R-project.org/.

Øien, N. 1995. Norwegian independent line transect survey 1995. Interne notat, nr. 8-1995, Havforskningsinstituttet.

Pearson, K. 1990. On the criterion that a given system of deviations from the probable in the case of a correlated system of variables is such that it can be reasonably supposed to have arisen from random sampling, Philosophical Magazine. Series 5. 50 (302); 157-175.

Pearl, J. 1988. Probabilistic Reasoning in Intelligent Systems: Networks of Plausible Inference. Morgan Kaufmann.

Pearl, J. 2009. Causality: Models, Reasoning and Inference. Cambridge University Press, 2nd edition.

Sakamoto, Y. and Akaike, H. 1978. Analysis of cross classified data by AIC. Annals Institute of Statistical Mathematics, 30, B, 185-197. Marine Ecology Progress Series. Vol.499; 203-216. https://doi.org/10.3554/meps10653.

Smith, A.N., Anderson, M.J., Millar, R.B., and Willis, T.J. 2014. Effects of marine reserves in the context of spatial and temporal variation: an analysis using Bayesian zero-inflated mixed models. Marine Ecology Progress Series 499; 203-216. https://doi.org/10.3554/meps10653.

Solvang, H., Haug, T., Gjøsæter, H., Bogstad, B., Hartvedt, S., Øien, N., and Lindstrøm, U. 2021. Distribution of rorquals and Atlantic cod in relation to their prey in the Norwegian high Arctic. Polar Biology 44. https://doi.org/10.1007/s00300-021-02835-2.

Solvang, H., Haug, T., and Øien, N. 2022. Recent trends in temporal and geographical variation in blubber thickness of common minke whales (Balaenoptera acutorostrata acutorostrata) in the Northeast Atlantic. NAMMCO Scientific Publications, 12. https://doi.org/10.7557/3.6308.

Solvang, H.K. and Planque, B. 2020. Estimation and classification of temporal trends to support integrated ecosystem assessment. ICES Journal of Marine Science 77; 2529-2540. https://doi.10.1093/icesjms/fsaa111.

Sugasawa, S., Nakagawa, T. Solvang, H.K., Subbey, S., and Alrabeei, S. 2022. Dynamic spatio-temporal zero-inflated Poisson models for predicting capelin distribution in the Barents Sea. Japanese Journal of Statistics and Data Science. https://doi.org/10.1007/s42081-022-00183-x.

The Institute of Statistical Mathematics. 2020. Package 'catdap' Categorical Data Analysis Program package, version 1.3.5.

Ver Hoef, J.M. and Jansen, J.K. 2007. Space-time zero-inflated count models of Harbor seals. Environmetrics 18; 697-712. https://doi.org/10.1002/env.873.